\documentclass[11pt]{report}
\renewcommand{\author}[1]{\subsubsection*{}}
\newcommand{\address}[1]{\subsubsection*{\it#1}}
\raggedright
\def\beginrefs{\begingroup\parindent=0pt\frenchspacing
     \parskip=1pt plus 1pt minus 1pt\interlinepenalty=1000
     \pretolerance=10000\hyphenpenalty=10000\everypar={\hangindent=0.42in}}
\def\endrefs{\endgroup}
\begin{document}

\chapter*{\bf Quasi-Steady State Cosmology}
\author*{G.  Burbidge}

\setcounter{chapter}{1}

\address{University of California, San Diego
Center for Astrophysics and Space Sciences and
Department of Physics
La Jolla, CA 92093--0424}

\section{History}

Modern cosmology began with the realization that there were solutions
to Einstein's theory of gravity discovered by Friedmann and Lemaitre which
when combined with the redshift distance relation of Hubble and others could
be interpreted as showing that we live in an expanding universe.   It was then
only elementary logic to argue that if time reversal was applied, the universe
must originally have been so compact that we could talk of a beginning.
Lemaitre tried to describe this state as the ``Primeval Atom''.  For a decade
or so after the war, Fermi, Teller, Maria Mayer, Peierls, Gamow, Alpher,
Herman and others explored this dense configuration trying to make the
chemical elements from protons, neutrons and electrons.  They soon learned that
this was not possible because of the absence of stable masses of five
and eight,
but it was also realized that \underline{if} such an early stage had occured,
the universe would contain an expanding cloud of radiation which would
preserve its black body form as it expanded.    Gamow and his colleagues
were  particularly
intrigued by the physics of this hot fireball.   Dicke and his
colleagues in Princeton
rediscovered this idea in the 1960's and decided to try and detect the
radiation.  Penzias and Wilson found such a radiation field, and COBE has
demonstrated that it has a perfect black body form out to radio  wavelengths.
This history  together with the fact that the light isotopes D,
He$^3$ and He$^4$ in about the right amounts can be made in a hot big
bang if the parameters are chosen correctly, has led to the widely held
view, that the standard cosmology -- the hot big bang -- is correct.

What was not properly understood by those who first discussed the early
universe was that McKellar had already discovered the microwave radiation in
1941 (he obtained a temperature of 2.3$^\circ$ K, setting as a lower limit
1.8$^\circ$ K, and as an upper limit 3.4$^\circ$ K for the black-body
temperature) (McKellar 1941).  Thus it is wrong to argue
that Penzias and Wilson found serendipitously the radiation
that Gamow and his colleagues had predicted would be there, since Gamow at
least was aware of McKellar's results.   The
only reason why the
physicists decided to invoke a dense configuration in an early
universe was to find a
place with a plentiful source of neutrons.   Since the model failed
to explain the building
of nearly all of the chemical elements (which we now know following
Hoyle were made
in the stars) the model might well have been dropped, as it was by
Fermi and his
colleagues.  This is especially clear when it is also pointed out
that in the 1950's both
Bondi, Gold and Hoyle (1955) and independently, Burbidge (1958)
pointed out that if the
observed abundance of He was obtained by hydrogen burning in  stars,
there must have
been a phase in the history of the universe when the radiation
density was much higher
than the energy density of starlight today.  The very striking fact
is that if we suppose
that if $\rho$ is the density of baryonic  matter in the universe, with
a value of
about 3 x 10$^{-31}$ gm cm$^{-3}$, and that the He/H ratio by mass in it is
0.244, then the energy which must have been released in producing He is
4.39 x 10$^{-13}$ erg cm$^{-3}$.  If this energy is thermalized, the black
body temperature turns out to be T = 2.76$^\circ$ K.   This value is
astonishingly close to the value of 2.73$^\circ$ K observed by COBE.

This simple agreement of two measured quantities makes no allowance for the
expansion of the universe that must necessarily have taken place during the
production of helium, which would act to reduce the temperature.   However,
it does show that unless it is dismissed as a coincidence which all big bang
believers must do, there  is likely to be an explanation of the microwave
radiation in terms of straight forward astrophysics involving hydrogen burning
in stars.

This line of reasoning completely refutes the
popular view that the discovery of the microwave radiation is proof that a big
bang occurred.
The usual rebuttal to this argument is that it is the blackbody nature of the
radiation that is important, not the value of the temperature, and that in any
alternative scheme it is the thermalization process of the radiation that is
the weak link.
The counter to that  is that in the standard model generation of the
black body radiation is traced to the decay of the false vacuum energy in the
inflationary scenario.   This, however, requires a gross extrapolation
beyond known physics.

\section{The Theory of the Quasi-Steady State Cosmology}

This theory was developed starting in the early 1990's (Hoyle,
Burbidge \& Narlikar
1993; 1994 a,b,; 1995; 2000).   The basic theory is the Machian theory of
gravity first proposed by Hoyle and Narlikar (1964, 1966b) in which the origin
of inertia is linked with a long range scalar interaction between matter and
matter.   Specifically, the theory is derivable from an action
principle with the
simple action:

\begin{equation}
A =  -\sum_{a} \int m_ads_a~~~ ,
\end{equation}

where the summation is over the particles in the universe, labelled by $a$,
the mass of the $a$th particle being $m_a$.  The integral is over the
world line of the particle, $ds_a$, representing the element of
proper time of the
$a$th particle.

The mass itself arises from interaction with other particles.  Thus the mass of
particle $a$ at point A on its word line arises from all other
particles $b$ in the
universe:

\begin{equation}
m_a =  \sum_{b\neq a} m^{(b)} (A)  ,
\end{equation}

where $m^{(b)}$ (X) is the contribution of inertial mass from
particle $b$ to any
particle situated at a general spacetime point $X$.   The long range effect is
Machian in nature and is communicated by the scalar mass function
$m^{(b)}$(X) which satisfies the conformally invariant wave equation

\begin{equation}
m^{(b)} + \frac{1}{6} Rm^{(b)} + [m^{(b)}]^3 = N^{(b)}~~  .
\end{equation}

Here the wave operator is with respect to the general spacetime point $X$.  $R$
is the scalar curvature of spacetime and the right hand side gives the number
density of particle $b$.   The field equations are obtained by
varying the action
with respect to the spacetime metric $g_{ik}$.    The important point
to note is
that the above formalism is conformally invariant.  In particular,
one can choose
a conformal frame in which the particle masses are constant.   If the constant
mass is denoted by $m_p$, the field equations reduce to

\begin{equation}
R^{ik} - \frac{1}{2} g^{ik} R + \lambda g^{ik} = - \frac{8 \pi G}{c^4}
[T^{ik} - \frac{2}{3} (c^i c^k - \frac{1}{4} g^{ik} c^l c_l)]~~~,
\end{equation}

where $c$ is a scalar
field which  arises explicitly from the ends of broken world lines,
that is when
there is creation (or, annihilation) of particles in the universe.  Thus the
divergence of the matter tensor $T^{ik}$ need not always be zero, as the
creation or annihilation of particles is compensated by the non-zero
divergence of
the c-field tensor in Equ.(4).   The quantities $G$ (the
gravitational constant) and
$\lambda$ (the cosmological constant) are related to the large scale
distribution
of particles in the universe.  Thus,

\begin{equation}
G = \frac{3 hc}{4 \pi rm^2_p}~~,  \lambda = - \frac{3}{N^2 m^2_p}~~,
\end{equation}

$N$ being the number of particles within the cosmic horizon.

Note that the signs of the various constants are determined by the theory
and not put in by hand.  For example, the constant of gravitation is
positive, the
cosmological constant negative and the coupling of the $c$-field
energy tensor to
spacetime is negative.

The action principle tells us that matter creation is possible at a
given spacetime
point provided the ambient $c$-field satisfies the equality $c = m_p$ at that
point.  In normal circumstances, the background level of the $c$-field will be
below this level.   However, in the strong gravity obtaining in the
neighborhood of
compact massive objects the value of the field can be locally raised.  This
leads to creation of matter along with the creation of negative
$c$-field energy.
The latter also has negative stresses which have the effect of blowing the
spacetime outwards (as in an inflationary model) with the result that
the created
matter is thrown out in an explosion.

We shall refer to such pockets of creation as minibangs or
mini-creation events.
A spherical (Schwarzchild type) compact matter distribution will lead to a
spherically symmetric explosion whereas an axi-symmetric (Kerr type)
distribution would lead to jet like ejection along the symmetric
axis.  Because of
the conservation of angular momentum of a collapsing object, it is
expected that
the latter situation will in general be more likely.

In either case, however, the minibang is nonsingular.   There is no state of
infinite curvature and terminating worldlines, as in the standard big
bang, nor is
there a black hole type horizon.  The latter because the presence of
the $c$-field causes the collapsing object to bounce outside the event horizon.

The feedback of such minibangs on the spacetime as a whole is to make it
expand.  In a completely steady situation, the spacetime will be  that given by
the deSitter metric.  However, the creation activity passes through
epochs of ups
and downs with the result that the spacetime also shows an
oscillation about the
long term steady state.  Sachs et al. (1996) have computed the simplest such
solution with the line element given by

\begin{equation}
ds^2 = c^2 dt^2 - S^2(t)[dr^2 +r^2 (d\theta^2 + sin^2 \theta d\phi^2)]~~,
\end{equation}

where $c$ stands for the speed of light and the scale factor is given by

\begin{equation}
S(t) = e^{t/P} [1 + n cos \frac{2 \pi r(t)}{Q}]~~.
\end{equation}

The constants $P$ and $Q$ are related to the constants in the field equations,
while $\tau$(t) is a function $\sim$ $t$ which is also determined by the field
equations.  For details see Sachs et al. (op.ci.).   The parameter $n$ may be
taken positive and is less than unity.  Thus the scale factor never
becomes zero:
the cosmological solution is without a spacetime singularity.
\vspace*{8pt}

{\bf Explosive Cosmogony}
\vspace*{8pt}

We have just pointed out that the the theory requires that creation
takes place in
the vicinity of already present massive objects.  This means that matter will
appear in the universe in the form of compact objects ejected from already
existing massive objects, i.e. the nuclei of galaxies.  Thus the
theory predicts that
galaxies beget galaxies,  most frequently when the universe is close to its
minimum phase, and less frequently as the universe expands in the
cycle.  In this
theory all of the creation takes place in regions of very high density.

\section{The Observed Properties of the Universe as they are Understood
in a (a)  Big Bang Model and (b) the QSSC}

\subsection{The Expansion of the Universe}

This was the first discovery which
led to the development of modern cosmology.    Since Einstein's theory
allows expanding (or contracting) solutions,   this discovery was
clearly a triumph
for general relativity.  If we consider the reversal of the time
axis this leads
immediately to the concept of a hot dense beginning phase at t = 0.    The
difficulty comes when we have to discuss the physics of the very 
early universe.

As the universe shrinks the radiation energy begins to dominate and
ultimately the matter is broken down into quarks.   We now
move out of the realm of known physics.  A further contraction by a factor of
about 10$^{10}$ is
invoked leading to what is called a ``phase transition" in which everything is
converted into a new kind of so-called scalar particle.   These
scalar particles are
supposed to interact together to produce what is described as a
``false vacuum''
maintaining positive energy at all costs.  This false vacuum consumes
space-time
in a process of deflation -- this is the inflation epoch of Guth and
Linde when time
is reversed.  The consuming of spacetime leads to what?   To a quantum
transition to somewhere else!   This gross extrapolation is in part
because there is
strong resistance  to the idea (of Dirac) of particles of negative
energy.  While the
energetics are still outside the realm of known physics, the existence of a
negative energy field will permit entirely new positives to form with the new
positives compensating those of negative energy with what we can refer to as
creation events in which energy is conserved.

This is the approach taken in the case of the QSSC.   It is derived
from the introduction of the C-field (Hoyle \& Narlikar 1964) which involves a
modification
of the theory of general relativity which only comes into play in very strong
gravitational flields.   Admittedly this is a classical field theory
which has not been
quantized.   At the same time we avoid the untestable aspects of the
theory close to  t = 0
where not only creation of matter and energy must be invoked, but
also the creation
of the laws of physics, which are assumed to be immutable,  and God
given.   Since in the QSSC there is no beginning, and creation takes place
in the nuclei of galaxies, observational tests of this theory are much easier
to make, since they continue to occur at all epochs.

\subsection{The Chemical Composition of the Universe}

In both the big
bang cosmology and the QSSC it is accepted that all of the chemical
elements heavier than $^7$Li have been synthesized in stars.  The
difference  between them lies in their different views of the theories in the
origin
of the lightest isotopes $^2$D, $^3$He, $^4$He and $^7$Li.

As far as the hot big bang is concerned Gamow and his colleagues
originally showed that these isotopes could be synthesized in an early
universe \underline{provided} that an appropriate value of
$\rho_b$/$\rho_r$ was chosen (Alpher \& Herman 1950).

The original intention was to choose a value of this ratio which would lead to
a value of the $^4$He abundance close to the observed value (in the
1950's) of Y = 0.25.
Thus they put

\begin{equation}
\rho_b = 1.70 \times  10^{-2} t^{-\frac{3}{2}} gm~cm^{-3}
\end{equation}

corresponding to $\rho_b$/$\rho_r$ = 1.4 $\times$ 10$^{-10}$.

For the modern calculations $\rho_b$/$\rho_r$ = 4.5 $\times$ 10$^{-10}$,
and it has been shown that with this value good agreement can be
reached between theory and observation for the relative abundances of
$^2$D,
$^3$He, $^4$He and $^7$Li.   The most critical isotopes are $^4$He,
where observationally Y = 0.24,  and $^2$D where $^2$D/H = 2
$\times$ 10$^{-5}$.   It is clearly a success for the hot big bang theory that
the choice of $\rho_b$/$\rho_r$ which gives the observed ratio $^4$He/(H
+ $^4$He), will also give the observed ratio $^2$D/H, but it must not be
forgotten  that the initial choice of a value of $\rho_b$/$\rho_r$ is
entirely \underline{ad hoc}.

In the QSSC the situation is very different, since the lightest isotopes must
have been synthesized  in stars.    The
very striking observational fact (devoid of all theory) that if the $^4$He
abundance in the matter
\underline{known} to exist was synthesized from hydrogen in stars, the
energy released when thermalized will have a black body temperature very
close to that observed in the CMB.  While this simple agreement of two
measured quantities makes no allowance for the expansion that has taken
place, it  is a strong observational argument for QSSC.   Burbidge
and Hoyle (1998) have shown that the other isotopes $^2$D, $^3$He and
$^7$Li, can be made either in flaring activity on the surfaces of stars, as is
known to occur in the sun and in other stars, or in incomplete hydrogen
burning in the interiors.   This will lead to variations in the abundances of
$^2$D etc.

\subsection{The Cosmic Background Radiation}

Since the observations by Penzias and Wilson (1965) the
interpretation of the
CMB as a remnant of the hot big bang universe has been  continuously
touted as the strongest evidence for this cosmological model.   This has
occurred despite the fact that its origin was not predicted by Gamow and then
found, as is often stated, since the radiation had already been discovered by
McKellar  \underline{before} Gamow's ``prediction'', and also
despite the fact that in the BB cosmology the temperature cannot be
predicted (cf Turner 1993).  The reasons for this situation are largely
attributable to the enthusiasm and belief in the theory by the big bang
cosmologists, and their unscrupulous shading of the historical record.
What the theory did predict is that the relict radiation
would have a perfect black body form, and the results from COBE  provide
strong positive evidence  for this model.

Development of theories for the formation of galaxies in the early universe
requires the invoking of initial density fluctuations in the matter.  These in
turn lead to fluctuations in the background radiation.   The fluctuation
spectrum which is expected depends on the parameters which are
chosen to obtain the observed large scale structure of visible galaxies.   A
well defined set of peaks is predicted for many classes of model but not all.
The favored set of models require the assumption of inflation,
and the presence of non-baryonic cold dark matter (CDM).  Making all of
these assumptions  it is possible to predict the relative positions and
amplitudes of an harmonic series  of acoustic peaks in the angular power
spectrum as a function of $\Omega_{tot}$, $\Omega_b$,
$\Omega_c$ and n$_s$.

Recent observations made by BOOMERANG and DASI (Netterfield et al.
2001; Pryke et al. 2001) have  led to the discovery of several maxima and
minima in the angular power spectrum which have led the authors to the
conclusions that they are detecting fluctuations which arise in the adiabatic
inflationary models and they can place limits on $\Omega_{tot}$,
$\Omega_b$, $\Omega_c$ and n$_e$.

How is the microwave background radiation understood in the QSSC?
It must have been generated by large numbers of discrete sources in
which mini-bangs occured and matter was created, over many cycles  of
oscillations as the universe slowly expands.  The basic energy generation
mechanism is the burning of hydrogen in massive stars which gives rise to
ultraviolet photons.  The photons are then degraded in energy by scattering,
absorption, and re-emission by dust particles,  reachng equilibrium at the CMB
observed temperature.  There are two questions:

\begin{enumerate}
\item[(a)] Can the radiation be thermalized to give rise to a highly isotropic
CMB with an almost exact black body form (at least out to radio wavelengths
of $\sim$10 cm)?

\item[(b)] Can we explain the angular fluctuations currently being found?
\end{enumerate}

HBN (2000 and earlier references) have given a detailed answer to (a).  The
scale factor for k = 0 is given by

\begin{equation}
S (t) = exp~ (\frac{t}{P})~[1 + \eta~cos~ (t)]
\end{equation}

where 0 $< \eta <$ 1, so that S oscillates between two finite values and
$\tau$(t) is almost like $t$ during most of the oscillatory cycle, 
differing from
it mostly during the stage when $S$ is close to the minimum value.   The
period of oscillation $Q$ is small compared to $P$.    The QSSC is therefore
characterized by the following parameters: $P, Q, \eta$ and $z_{max}$, the
maximum redshift seen by the present observer in the current cycle.  Sachs
et al. (1996) took $P = 20Q, Q = 4.4 \times 10^{10}$ yrs, $\eta$ = 0.8,
$z_{max}$ = 5, as an indicative set of values.  Thus the
QSSC oscillations are finite with the maximum redshift observable in the
present cycle at $\sim$5--6, and each cycle is matter-dominated.   The
radiation background is however, maintained from one cycle to next.
Thus from the minimum scale phase of one cycle to next, its energy density is
expected to fall by a factor exp $(-4 Q/P)$.  This drop is made up by the
thermalization of starlight produced during the cycle.   Thus if 
$\theta$ is the
energy density of  starlight generated in a     cycle
and $u_{max}$ is the energy density of the CMB at the start of a cycle, then
$\epsilon~^\sim_=~4~u_{max}~Q/P$.   If the cycle minimum occured at
redshift $z_{max}$, then the present CMB energy density would be
$P \epsilon/4Q(1+z_{max})^4$.     Substituting the values of $\theta$, $P$,
$z_{max}$ and Q chosen above we can estimate the present day energy
density of CMB and the result agrees well with the observed value of $\sim$
$4 \times 10^{-13} erg~cm^{-3}$ corresponding to temperature $\sim$
2.7K.

How is the starlight thermalized?   There is
good laboratory evidence that the cooling of metallic vapours including carbon
produces whisker-like particles of lengths $\sim$ 0.5 -- 1.0 mm, which convert
optical radiation into millimetre radiation.   Such whiskers will 
typically form in
the neighborhood of supernovae which eject metals, and are subsequently
pushed out of the galaxy through pressures of shock waves.  It has been
shown (HBN 2000) that a density of $\sim$ 10$^{-35}$ g
cm$^{-3}$ of such whiskers close to the minimum of the oscillatory phase is
sufficient for thermalization of starlight.  Narlikar et al. (1997) 
have discussed
evidence for such whiskers in different astrophysical settings.

While the thermalized radiation from previous cycles will be very smoothly
distributed, a tiny fraction ($\sim 10^{-5}$) will reflect anisotropies on the
scales of rich clusters of galaxies in the present cycle.  The 
angular scales for
this anisotropy will be of the order $\sim$ 1/100, --1/250 for superclusters
corresponding to values $l$ values $\sim$ 100--200.

Thus in QSSC it is the localized effects of radiation associated with 
individual
clusters or superclusters of galaxies in which mini-bangs have occurred
that are responsible for the observed fluctuations.   It has recently been
shown (Narlikar et al. 2001) that rich clusters on the scale of 5--10 Mpc at
the redshift epoch close to z = 5  can generate the first major peak in the
fluctuation spectrum observed by BOOMERANG and MAXIMA.   While no
detailed modelling has yet been made it is very likely that the smaller peaks
at larger values of $l$ will arise from the effects of dust generated and
expelled from individual galaxies and small groups.

\subsection{Redshifts of QSOs and Explosive Events in Active Galactic Nuclei}

The existence of
a class of objects which have redshifts not largely due to the cosmic
expansion was not predicted either in the hot big bang cosmology or in
QSSC.  How is this phenomenon dealt with in each hypothesis?

As far as that big bang model is concerned its supporters are in complete
denial.   They never mention the  observational evidence, do not
allow observers who would like to report such evidence any opportunity
to do this in cosmology conferences, argue against its publication, and if
forced to comment on the data, simply argue that they are wrong.

The evidence is strong (for a review see Burbidge 2001), and so it must be
explained.    It
suggests that QSOs and related condensed objects with anomalous
redshifts are ejected from the centers of active galaxies at all cosmological
redshifts and populate the general field.  Thus QSOs cannot be used for
classical cosmological investigations .

The explosive cosmogony associated with the QSSC is discussed in HBN
(2000).   It is based on the idea that new matter and energy is generated close
to the massive centers of galaxies, and it then moves outward, i.e. this is  an
explosive cosmogony rather than the cosmogony required in the big bang
where it is supposed   that all condensed objects are formed early in the
expansion as a result of gravitational instability and collapse of 
initial density
fluctuations.  It is therefore a \underline{prediction} of the QSSC 
that ejection
processes involving massive condensed objects, gas moving at high speeds,
and large fluxes of radiation will be commonplace.  Thus the  existence of
active galactic nuclei, radio sources, expanding shells and even ejected
QSOs all can be considered to be evidence in favor of this idea.   However the
anomalous redshifts have not been really understood in either cosmological
theory.   Within the framework of the known laws of physics redshifts can
only arise as expansion shifts, Doppler shifts (both redward and blueward),
gravitational shifts, or shifts due to atomic transitions involving 
particles with
masses different from electron masses.   Hoyle \& Burbidge (1996) have
tried to understand how such shifts could occur when the masses of 
the particles
vary, but we have no theory which can explain the remarkable peaks and
periodicity  in the distribution of the redshifts of these objects which have
been found over the last 30 years (cf Burbidge \& Napier 2001).

\subsection{Galaxies and Large Scale Structure}

Within the framework of
the conventional cosmology there will be no gravitationally stable
configurations formed in the expanding universe unless it is
\underline{assumed} that initial density fluctuations are present, and also
that there is a major component of mass energy in the form of non-baryonic
matter.
Given that all of these assumptions are made and also that inflation has taken
place, it is possible to carry out numerical simulations of what we would
expect in the form of galaxies, with a dominant component of
non-baryonic dark matter.

Comparisons are also attempted between the predictions of numerically
simulated models of large scale structure and what can be deduced about
the properties of matter from the absorption spectra (in particular the
Ly$\alpha$\\ forest) of  high  redshift QSOs.   All of this latter 
depends on yet
another assumption,  that the redshifts of the QSOs are completely of
cosmological origin and that the absorption is due to intervening gas.   These
assumptions are generally believed but they may not be correct.

How are galaxies and large scale structures to be understood in terms of the
framework of the QSSC?    There is no early universe and
galaxies are not made as a result of gravitational collapse and
condensation.  Instead as was described in the previous section they arise as
a result of creation in the vicinity of massive objects (the nuclei of existing
galaxies) and are ejected at least in part as coherent objects.   We do not
yet understand the detailed physics of these processes, but there is
extensive direct observational evidence at all redshifts that this mechanism
of explosive cosmogony is at work.   As was pointed out earlier this process
of galaxies begetting galaxies is expected to take place throughout the
oscillatory cycles in the QSSC, but most of the new galaxies will be born
near the minima of the oscillations, and it is those events which actually are
responsible for the universe re-expanding  without reaching the extremely
high density of the early universe in the big bang.

\subsection{The m-z Relation}

Many attempts have been made to measure
the shape of the m-z relation as a way to investigate cosmological models.
In terms of the scale factor $S(t)$, the Hubble constant H and the deceleration
parameter $q(t)$ are defined as

\begin{equation}
H (t) = \dot{S}(t)/S(t)~~ .
\end{equation}

and

\begin{equation}
q(t) = - \frac{1}{H^2}~ [\ddot{S}(t) / S (t) ]~~.
\end{equation}

For all Friedmann models without the cosmological constant,  $q_o$ is expected
to be positive, corresponding to a decelerating universe.

On the other hand in the classical steady model, $q_o = -1$, and in the
quasi-steady state cosmology (QSSC) $q_o$ will also be negative, corrsponding
to an accelerating universe.

To obtain an accelerating universe within the framework of the Friedmann
model it is necessary to insert a positive  cosmological constant.

Recent determinations of the second order term using supernovae of Type Ia as
standard candles have been made by Perlmutter et al. (1999) and Riess et al.
(1998, 2001).
They have shown that the universe is accelerating.   This then is a 
result which
was \underline{predicted} by the steady state theory and the QSSC,  and it
has been shown by Banerjee et al. (2000) that a good fit can be made to the
observed data using QSSC.

Thus this observational evidence confirming a real prediction is 
clearly evidence
in favor of QSSC.  The heavy prejudice among cosmologists against the QSSC
is clearly illustrated by the way that the hot big bang supporters have
interpreted these results.   All of them including the observers who made the
supernova observations,  have claimed that these results can only be explained
by the introduction of a negative cosmological constant $\lambda$, leading to
their belief in what they call ``dark energy'' or ``quintessence''.

\subsection{The $\Theta - z$ Relation}

Recently, $\Theta - z$ relation has received special attention in the 
context of
ultracompact radio sources.  Kellermann (1993), Gurvits (1994) and Jackson
and Dodgson (1997) have used the fact that an ultracompact VLBI-detected
source, being deeply embedded in a radiosource will not be susceptible to
evolutionary effects on its size arising from the changes in the intergalactic
medium.  Using such a population of high redshift ($z > 0$) objects they were
able to argue that the dependence of angular size $\Theta$ on redshift $z$ can
be used to constrainthe cosmological models.  While Kellermann (op.cit.) found
the Einstein-de Sitter model (the standard $\Omega = 1$ model) consistent
with his data, Jackson and Dodgson, with their increased database found
that  the
model gives a marginally good fit.  They found that models with a large
negative cosmological constant give a better fit to the data.

Against this background, Banerjee and the Narlikar  (1997) have found that
the QSSC model gives a better (and very
good) fit to the $\Theta - z$ data.  In particular, the flattening of 
the curve at
large redshifts is in conformity with the data.

\section{References}

\beginrefs
Alpher, R.A., and Herman, R. 1950, Rev. Mod. Phys., 22, 153

Banerjee, S. and Narlikar, J.V. 1997, IUCCA preprint

Banerjee, S. et al. 2000, Astron. J., 119, 2583

Bondi, H., Gold, T., and Hoyle, F. 1955, Observatory, 75, 80

Burbidge, G. 1958, PASP, 70, 83

Burbidge, G. and Hoyle, F. 1998, ApJ., 509, L1

Burbidge, G. and Napier, W. 2001, Astron. J., 121, 21

Burbidge, G. 2001, PASP, August

Gurvits, L.I. 1994, ApJ, 425, 442

Hoyle, F., and Burbidge, G. 1996, A\&A, 309, 335

Hoyle, F., Burbidge, G. and Narlikar, J.V. 1993, ApJ, 410, 437

Hoyle, F., Burbidge, G. and Narlikar, J.V. 1994a, A\&A, 289, 729

Hoyle, F., Burbidge, G. and Narlikar, J.V. 1994b, MNRAS, 267, 1007

Hoyle, F., Burbidge, G. and Narlikar, J.V. 1995, Proc. Roy. Soc. A, 448, 191

Hoyle, F., Burbidge, G. and Narlikar, J.V. 2000, A Different Approach to
Cosmology (Cambridge University Press)

Hoyle, F. and Narlikar, J.V. 1964, Proc. Roy. Soc. A, 290, 191

Hoyle, F. and Narlikar, J.V. 1966a, Proc. Roy. Soc. A, 294, 162

Hoyle, F. and Narlikar, J.V. 1966b, Proc. Roy. Soc. A, 294, 138

Jackson, J. and Dodgson, M. 1997, MNRAS, 285, 806

Kellerman, K. 1993, Nature, 361, 134

McKellar, A. 1941, Pub. Dom. Astrophs. Obs., 7, 251

Narlikar, J.V. et al. 1997, Int. J. Mod. Phys. D 6, 125

Narlikar, J.V. et al. 2001, submitted to ApJ

Netterfield, C. et al. 2001, preprint

Penzias, A., and Wilson, R.W. 1965, ApJ, 142, 419

Perlmutter, S. et al. 1999, ApJ, 517, 565

Pryke, A. et al. 2001, preprint

Riess, A. et al. 1998, Astron.J., 116, 1009

Riess, A. et al. 2001, preprint

Sachs, R., Narlikar, J.V. and Hoyle, F. 1996, A\&A, 313, 703

Turner, M. 1993, Science, 262, 861
\endrefs

\end{document}